\documentclass[apj]{emulateapj}\slugcomment{{\sc Accepted to ApJ:} 2013 April 21}
\shorttitle{Polarized Foregrounds}
\shortauthors{Moore, et al.}

\usepackage{amsmath}
\usepackage{amsfonts}

\usepackage{natbib}
\newcommand{\nar}{NAR}
\newcommand{\pasa}{PASA}

\newcommand{\V}{\mathcal{V}}
\newcommand{\B}[1]{\mathbf{#1}}
\newcommand{\uv}{$u$-$v$\ }

\begin{document}

\title{\uppercase{The Effects of Polarized Foregrounds on} 21 \lowercase{cm} \uppercase{Epoch of Reionization Power Spectrum Measurements}}
\author{David F. Moore\altaffilmark{1},
        James E. Aguirre\altaffilmark{1},
        Aaron R. Parsons\altaffilmark{2},
        Daniel C. Jacobs\altaffilmark{3}, and
        Jonathan C. Pober\altaffilmark{2}
        }
\altaffiltext{1}{Dept. of Physics and Astronomy, U. Pennsylvania, Philadelphia, PA, damo@sas.upenn.edu}
\altaffiltext{2}{Astronomy Dept. U. California, Berkeley, CA}
\altaffiltext{3}{School of Earth and Space Exploration, Arizona State U., Tempe, AZ}

\begin{abstract}
    Experiments aimed at detecting highly-redshifted 21 centimeter emission from the Epoch of Reionization (EoR) are plagued  by the contamination of foreground emission. A potentially important source of contaminating foregrounds may be Faraday-rotated, polarized emission, which leaks into the estimate of the intrinsically unpolarized EoR signal. While these foregrounds' intrinsic polarization may not be problematic, the spectral structure introduced by the Faraday rotation could be. To better understand and characterize these effects, we present a simulation of the polarized sky between 120 and 180 MHz. We compute a single visibility, and estimate the three-dimensional power spectrum from that visibility using the delay spectrum approach presented in \citet{DelaySpectrum}. Using the Donald C. Backer Precision Array to Probe the Epoch of Reionization (PAPER) as an example instrument, we show the expected leakage into the unpolarized power spectrum to be several orders of magnitude above the expected 21cm EoR signal. 
\end{abstract}


\section{Introduction}

A significant amount of thought has gone into the problem of foreground removal allowing for the detection of the power spectrum of neutral hydrogen during the Epoch of Reionization (EoR) \citep[e.g.]{Bowman2009,Morales2006,Liu2009,LiuTegmark2011,Dillon2012}. Essential to nearly all these techniques is the spectral smoothness of these foregrounds, which allows for a separation in $k$ space of the foreground emission and the signal from the EoR. Various mechanisms will leak polarized emission into the best estimate of  the EoR signal, which then may introduce spectral structure to an EoR experiment's measurement. 

\begin{figure}
    \plotone{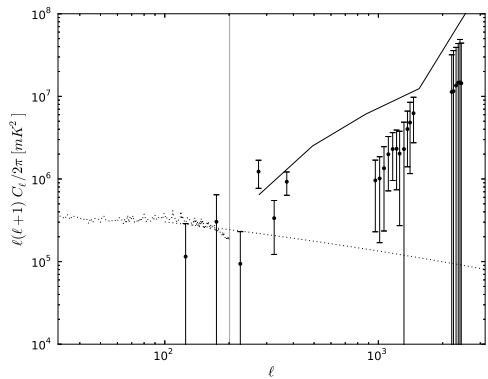}
    \caption{\label{fig:measurements} Recent estimates of low-frequency polarized power spectra. The thick black points
    with error bars show the \citet{Bernardi2010} measurements of a field around 3C196, and the solid line shows the
    upper limit of \citet{Pen2009}. The Haslam map at 408 MHz \citep{Haslam}, scaled by a polarization fraction of
    0.3\%, is shown by thin points, and a power-law extrapolation is shown with a dotted line above $\ell =200$. This
    fraction was chosen to agree with the low-$\ell$ points in the Bernardi measurement. At high-$\ell$, the upper
    limits do not constrain the level of polarized emission. The grey vertical line shows the $\ell$ mode sampled by the simulation in this paper, as a point of reference.}
\end{figure}

By contrast to unpolarized foregrounds, comparatively little work has addressed the problem of detecting and removing
polarized foregrounds. \citet{Pen2009} provided some of the first relevant upper limits at small angular scales using
the Giant Metre-Wave Telescope (GMRT). They found, for spherical harmonic multipoles $200 \le \ell \le 5000$, an angular
power spectrum upper limit of around $C_\ell \lesssim 100$~mK$^2$. More recent work by \citet{Bernardi2010} detected
polarized power at the same level at $\ell \lesssim 1000$ using the Westerbork Synthesis Radio Telescope (WSRT), with no
significant detection above $\ell$ of 1000. \citet{Bernardi2010} did not detect emission directly attributable to
polarized point sources. \citet{Pen2009} present their upper limit in terms of the three-dimensional power spectrum, but
\citet{Bernardi2010} only present a $C_\ell$ spectrum which has been integrated along the frequency direction. Figure
\ref{fig:measurements} gives a summary of the low-frequency measurements of polarized power spectra. It is clear that
the angular power spectrum of the unpolarized sky must be scaled by a mean polarization fraction of about $0.3\%$ to
agree with the \citet{Bernardi2010} measurements. This requires a significant degree of depolarization of synchrotron
emission from ordered magnetic fields.  

\citet{Jelic2010} attempted  to further constrain the problem by fully simulating  a full-Stokes realization of galactic
synchrotron emission over a $10^\circ \times 10^\circ$ field of view. They present a realistic spectrum of the mean
temperature of polarized emission, but do not extend their analysis into the power spectrum. Their analysis also
predicts a polarization fraction from diffuse emission much higher than the limited measurements available. \citet{Geil2011} also investigate the issue, proposing the use of the RMCLEAN algorithm \citep{Heald2009} to mitigate the effects of these polarized foregrounds. While these papers provide detailed descriptions of the polarized sky and removal strategies in the image plane, they provide little discussion of the line-of-sight direction.

This paper aims to steer the discussion of polarized foregrounds towards the \emph{terra incognita} of the third, frequency dimension. We begin in section \ref{sec:prelim} reviewing the basics of polarized interferometry, and the delay spectrum approach to estimating the power spectrum, presented in \citet{DelaySpectrum}. We discuss the relevance of polarized foregrounds to measuring the 21cm power spectrum. In section \ref{sec:SIM}, we discuss the design and implementation of a simulation of the polarized sky and its results. Finally, in section \ref{sec:remove} we briefly discuss prospects of polarized source removal and leakage mitigation.

Throughout the paper, we will use the Donald C. Backer Precision Array to Probe the Epoch of Reionization (PAPER)\citep{PGB8} as a model instrument. The results are not specific to that instrument, nor are they specific to the Delay Spectrum analysis used in this paper \citep{DelaySpectrum}. Any 21cm EoR power spectrum detection experiment with linear feeds, including the Murchison Widefield Array \citep{MWA} or the  Low Frequency ARray \citep{LOFAR} could fall subject to the leakages described here without a perfect calibration.

\section{Preliminaries}\label{sec:prelim}

\subsection{Definition of Stokes Visibilities}\label{sec:stokes}

Two prominent ways in which polarized sky emission can leak into an interferometeric estimate of Stokes $I$ are leakages due to non-orthogonal and rotated feeds and beam ellipticity --- an asymmetry in the two linear polarizations of a primary beam which causes unpolarized signals to appear polarized, and vice versa. The first is a well-understood question, discussed in length in the series of papers by \citet{HBS1}. This type of leakage can be corrected by the proper linear combination of visibilities. Hence, we will focus on the latter issue. To begin, we will examine the contents of an interferometric spectrum, and relate them to the intrinsic Stokes parameters.

As a reminder, we present the measurement equation for an interferometer in the flat-sky limit:
\begin{equation}
    \V_{ab}(u,v,\nu) =  \int A_{ab}(l,m)I_{ab}(l,m)e^{-2\pi i (uv + lm)}\ dldm.
    \label{eq:vis_def}
\end{equation}
Here the polarization indices $a,b$ indicate the polarization state of the measurement. $A_{ab}$ is the primary beam, $l$ and $m$ are direction cosines of the celestial sphere (with their Fourier components $u$ and $v$), and $I_{ab}$ is the sky emission projected along the polarization state of the measurement. Henceforth, we will be writing all visibility equations in the flat-sky limit. We can do this without loss of generality since the polarization properties of a visibility are unaffected by this assumption.

It is worth noting that each linear feed measures a one-dimensional projection of the incident electric field. This causes correlations with different feed orientations to contain information about different polarization states of the incident radiation.  A convenient short-hand notation for the polarization content of each visibility is 

\begin{equation}
    \begin{pmatrix}\V_{xx} & \V_{xy} \\ \V_{yx} & \V_{yy} \end{pmatrix}_{ij} =
        \B{J}_i \cdot 
        \begin{pmatrix} I+Q & U - iV \\ U + iV & I - Q \end{pmatrix} 
        \cdot \B{J}^\dagger_{j}
    \label{eq:vis_22}
\end{equation}

The Jones matrices $\B{J}_{i,j}$ \citep[see][]{HBS1} relate the sky emission ($I$, $Q$, $U$, and $V$) to an interferometeric measurement. Each antenna's Jones matrix is dependent on nearly all instrumental parameters, but for the purposes of this paper, we will investigate the effects of direction-dependent gains, or the primary beam. 

In the flat-sky approximation, an interferometer natively measures the two-dimensional spatial Fourier transform of the sky. Ideally, this would allow an observer to estimate the three-dimensional power spectrum by simply transforming the frequency, line of sight direction, and cross-multiplying measurements without imaging. Relaxing the imaging requirement provides an incentive to make estimates of the Stokes parameters --- $I$ for instance --- in the visibility domain, where the Stokes parameters are not defined. A na\"{i}ve addition between the linearly polarized, $xx$ and $yy$ visibilities should estimate the total power of sky emission, Stokes $I$. Similar operations can be performed for all polarization states. A sensible method of estimating the four Stokes parameters in the visibility domain is to add visibilities as images are typically added. Hence, we define Stokes visibilities (for linear feeds) as:

\begin{equation}
    \begin{pmatrix} \V_I \\ \V_Q \\ \V_U \\ \V_V \end{pmatrix} \equiv \frac{1}{2}
        \begin{pmatrix} 1 &  0 & 0 &  1 \\
                        1 &  0 & 0 & -1 \\
                        0 &  1 & 1 &  0 \\
                        0 & i & -i &  0 \end{pmatrix}
        \begin{pmatrix} \V_{xx} \\ \V_{xy} \\ \V_{yx} \\ \V_{yy} \end{pmatrix}.
    \label{eq:stokes_def}
\end{equation}

Explicitly writing the expression of $\V_I$ by substituting equation \ref{eq:vis_def} into equation \ref{eq:stokes_def}, we find that
\begin{align}
	\V_I = &\V_{xx} + \V_{yy} \nonumber \\
		= &\int \left(A_{xx} + A_{yy}\right) I e^{-2\pi i (lu +vm)}\ dldm \nonumber \\
		&+ \int \left(A_{xx} - A_{yy}\right)Q e^{-2\pi i (lu + vm)}\ dldm.
	\label{eq:visI}
\end{align}
Writing the visibility this way highlights one source of polarized leakage: that due to beam ellipticity. If the $xx$ and $yy$ beams are not equal, the last term of the equation \ref{eq:visI} would be non-zero. Polarized emission then enters the estimate of $I$, weighted by the differenced beam. This equation also points out the difference between Stokes parameters and the Stokes visibilities defined in equation \ref{eq:stokes_def}. They estimate their respective Stokes parameters, but still include terms from other polarization states. 
 
\begin{figure}
    \plotone{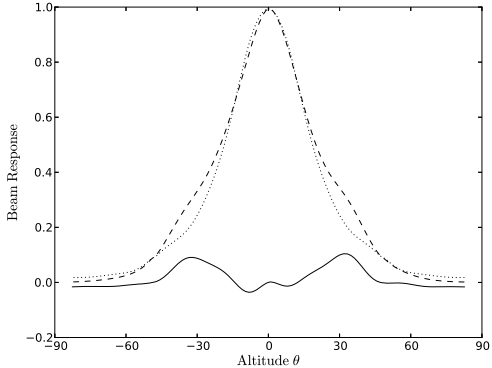}
    \caption{\label{fig:bm} Power vs. altitude through zenith for east-west slices of the $xx$ and $yy$ PAPER beams (dashed, dotted respectively). The solid line shows their difference. }
\end{figure}

Thus, if an instrument's linearly polarized beam is not symmetric under a $90^\circ$ rotation, all the intrinsic linear polarization will not cancel, corrupting the visibility estimate of Stokes parameters. Figure \ref{fig:bm} shows two slices through the model beam for PAPER, through constant azimuth and through zenith, and a cut through the same beam, rotated by $90 ^\circ$. We note that these slices are not identical, and show the differenced beam, which provides the mechanism for leakage. 

In an image-based analysis, these effects may be corrected by a simple re-weighting of the image, but these reconstructions are subject to the accuracy of the primary beam model. The results of this paper indicate the leakages due to the primary beam shown in figure \ref{fig:bm}, but could equally be applied to an image-based analysis whose primary beam is only known to the level of the differences in that figure.

Once more, we note that the Stokes visibilities are only the best guess at the true Stokes parameters. Without an exact characterization of the primary beam or dense \uv sampling, one does not have the ability to fully correct for the leakages mentioned. Hence, it is imperative to inspect the Stokes $I$ signal's corruption. 

\subsection{The Delay Spectrum Approach}\label{sub-sec:dsa}

The delay-spectrum approach to measuring the 21cm power spectrum \citep{DelaySpectrum} requires no imaging, again providing an incentive for adding raw visibilities. The delay-spectrum approach embraces the natural units of an interferometer by sampling the power spectrum in baseline and frequency dimensions, native to an interferometer. With this approach, each baseline can be individually transformed into an estimator of the power spectrum of the incident temperature. This method also prevents the small-scale structure introduced by Fourier transforming over gain calibration errors \citep{Morales2012}.

Perhaps the greatest advantage of the delay spectrum approach is its ability to isolate smooth-spectrum foregrounds on a
per-baseline basis. This relaxes the burden of isolating sources in the image plane, and allows for a more sparsely
sampled array (e.g. a redundantly sampled one, as in \citealt{PAPERSensitivity}). The 21cm EoR's extent to super-horizon delays depends precisely upon its spectral non-smoothness. It is imperative, then, to identify and characterize spectrally non-smooth foregrounds that may corrupt the EoR measurement.

The delay spectrum approach is particularly useful for simulating the effects of a systematic error on the power spectrum. Because each visibility estimates the power spectrum, a simulation needs only create one visibility, which encapsulates much of the power spectral information needed for analysis. The computational cost is additionally reduced by the ability to calculate a power spectrum by Fourier transforming only one visibility, rather than gridding several visibilities into a $(u,v,\nu)$ cube and Fourier transforming along each $u,v$ pixel.

It is important to note that the results of this paper are not limited to delay-spectrum-based analyses. All effects mentioned in this paper are native to the sky and its expected spectral structure. Any analysis technique will incur all the same issues; we use the delay spectrum as a convenient probe for these issues. 

We briefly review the definition of the delay transform as a Fourier transform over the frequency axis of a visibility:

\begin{equation}
	\tilde{\V}(\tau) \equiv \int_{\Delta\nu} \V(u,v,\nu) e^{-2\pi i \tau \nu} d\nu,
	\label{eq:dspec}
\end{equation}

where $\nu$ represents the frequency, $\tau$ is the delay or the Fourier transform pair to frequency, whose physical
meaning is given in \citet{DelaySpectrum}, and the tilde denotes a delay-transformed visibility. We treat this as an
estimator of the Fourier transform of the temperature field, squaring it to derive an estimator of the 3-D power
spectrum $P(\vec{k})$,

\begin{equation}
 	\tilde{\V}^2(\tau) = |\tilde{\V}(\tau)|^2 \propto P(\vec{k}).
	\label{eq:dspec_square}
\end{equation}

where $\vec{k}$ is the wavemode corresponding to the measurement. To better mimic actual measurements, we divide the band into ten, 6-MHz sub-bands, and perform the delay transform on each individual band. The sub-bandwidth is chosen to sample the maximum cosmological distance over which the 21cm signal is expected to be cotemporal \citep[e.g.]{Wyithe2004,Furlanetto2006}. Once these transforms are computed, we multiply each spectrum by the appropriate factors to obtain the power spectrum in units of temperature squared. We write the expression converting the squared, delay-transformed visibility $\tilde{\V}(\tau)$ into a ``unitless,'' cosmological power spectrum $\Delta^2(k)$ as 

\begin{equation}
	\tilde{\V}^2(\tau) \approx \left(\frac{2k_B}{\lambda^2}\right)^2 \frac{\Omega B}{X^2Y} 
		\frac{2\pi^2}{k^3}\Delta^2(k),
	\label{eq:unit_change}
\end{equation}

where $k_B$ is the Boltzmann constant, $\lambda$ is the observing wavelength, $\Omega$ is the solid angle of the primary
beam, $B$ is the observed bandwidth, and $X$ and $Y$ are cosmological scalars which convert observed angles and
frequencies in to $h\mathrm{Mpc}^{-1}$, appropriately derived from Equations 3 and 4 of \citet{FOB}.

\citet{DelaySpectrum} offers a much more detailed discussion of this approach, beyond the scope of this paper. 

\subsection{Sparse $u$-$v$ Sampling and Wide-Field Polarimetry}\label{sub-sec:wide-field}

Another advantage of the delay spectrum approach is that it relaxes the requirement of gridding in the \uv plane. Each baseline is assigned a position in the \uv plane \emph{ab initio}, and visibilities from similar baselines may be coherently added without imaging. This allows for sparse sampling in the \uv plane without damaging effects from side lobes or missing data, problems other methods may experience. Since the delay spectrum rotates a power-spectrum estimate into the native coordinate system of an interferometer, there are no inherently missing frequency-data. \citet{PAPERSensitivity} presents the sensitivity benefits of a sparse, redundant array configuration, but other techniques aim to uniformly sample the $(u,v,\nu)$ cube, in order to mitigate the systematic effects of computing a Fourier Transform across unevenly sampled data.

An obvious disadvantage of having sparsely-sampled data is poor imaging. Not only does sparse sampling provide a highly irregular synthesized beam, but it also limits the available information for a full reconstruction of the image. Without adjacent \uv samples, a full, accurate deconvolution by a wide beam simply has insufficient information. As we will see, the inability to correct for beam effects will provide a significant source of systematic error via polarized leakage.

By choosing to wield the full power of the delay spectrum approach and redundant sampling, an observer is forced to add
visibilities with no beam weighting. The beam information supplied by adjacent \uv samples simply does not exist, and
without transforming into the image plane is unrecoverable. Hence, the imperative to investigate the implications of a
lack of beam-weighting, the na\"{i}ve construction of the $I$ visibility, arises.

Together, redundant sampling and the delay spectrum approach give a 21cm EoR experiment incentive to add raw
visibilities, subjecting it to potential leakage. An elliptical primary beam gives a mechanism whereby polarized
emission can corrupt an estimate of the total power. To what degree does polarized emission corrupt an estimate of the
21cm EoR signal? We will begin answering this question by characterizing the spectral non-smoothness that will possibly
arise from the rotation measure structure of polarized leakages.

\subsection{Faraday Rotation}

Faraday Rotation affects the polarization properties of an electromagnetic wave traveling through a plasma containing a magnetic field \citep{RybickiAndLightman}. The circular polarization oriented in a right-handed fashion to the direction of the incident field will be slowed by the plasma. This causes a rotation of the E-field's polarization angle by 

\begin{equation}
    \Delta\phi = \lambda^2 \frac{e^3}{(m_ec^2)^2}\int B_{||}(s)n_e(s) ds \equiv \lambda^2 RM.
    \label{eq:RMdef}
\end{equation}

Equation \ref{eq:RMdef} defines the rotation measure ($RM$) by which we characterize this phase wrapping. Since the Stokes parameters characterize the square of the electric field, the phase of the polarization vector is shifted by twice the angle defined in equation \ref{eq:RMdef}. After a signal passes through a Faraday screen, we measure a rotated polarization angle,
\begin{equation}
    \left( Q + i U \right)_{meas} = \left( Q + i U \right)_{int} e^{2i RM\lambda^2},
    \label{eq:faraday_rot}
\end{equation}
where the subscript ``meas" denotes the measured $Q$ and $U$ measurements, and ``int" denotes the intrinsic $Q$ and $U$ signal, as would be measured behind the Faraday screen.

\section{Simulation}
\label{sec:SIM}

\subsection{Single-Source Power Spectrum}\label{sec:single_source}

We begin our investigation of the effects of polarized foregrounds on the 21cm EoR signature by examining the power spectrum of a single source at zenith, whose signal has the structure of a single $RM$. In doing so, we can develop an intuition for the rotation measures that affect cosmologically interesting $k$ modes of the power spectrum. By looking at what is effectively the impulse response of a Faraday screen on the power spectrum, it will be easier later to interpret a more complicated model.

\begin{figure}
    \plotone{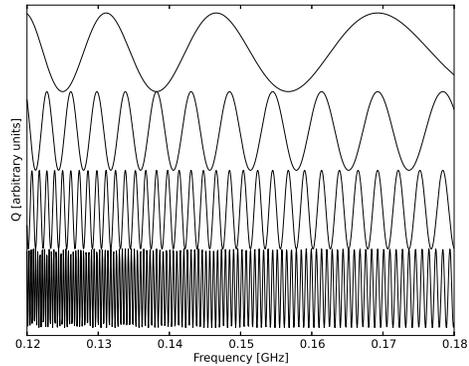}
    \caption{\label{fig:rm_spec} Real component of the visibility $\V = \exp\{-2RM\lambda^2\}$ for rotation measures 3, 10, 30 and 100 $m^{-2}$, from top to bottom. This plot is intended to demonstrate both the frequency-dependent phase wrapping and the frequency-dependent critical sampling of Faraday rotation. Notice how the 100 $m^{-2}$ component (located at the bottom) is critically sampled at high frequencies, but loses coherence at low frequencies.}
\end{figure}

Figure \ref{fig:rm_spec} shows the real part of the spectra of a few linearly polarized sources behind Faraday screens, $S(\nu) =
\exp\{-2iRM\lambda^2\}$. Each spectrum contains one source with one-Jansky of polarized flux, located at zenith (delay
of zero). Notice that at the highest rotation measure shown, the spectrum is not critically sampled at the lowest frequencies. This is due to the uneven sampling of $\lambda^2$ across the band: as $\Delta\lambda^2 \approx d\lambda^2/d\nu\ \Delta\nu \propto \Delta\nu/\nu^3$ increases, the sensitivity to large rotation measures decreases.

\begin{figure}
    \plotone{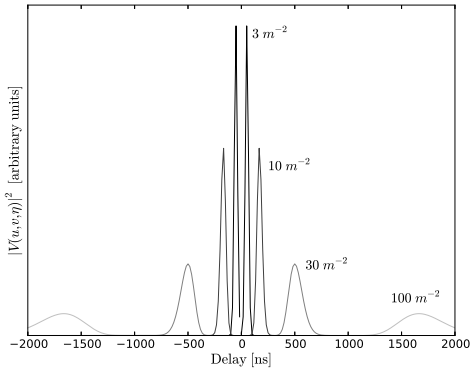}
    \caption{\label{fig:bh_pspec} Delay spectra of the Faraday-screened visibilities shown in figure \ref{fig:rm_spec}. This plot demonstrates the effect of a Faraday screen on the delay spectrum of a source. There are two effects: first, a Faraday screen widens the response of a spectrum in delay space, and second, the Faraday screen scatters power to high delay. This causes a potential problem as many foreground removal techniques require smooth-spectrum foregrounds. Note that higher rotation measures mimic higher delays.}
\end{figure}

\begin{figure}
    \plotone{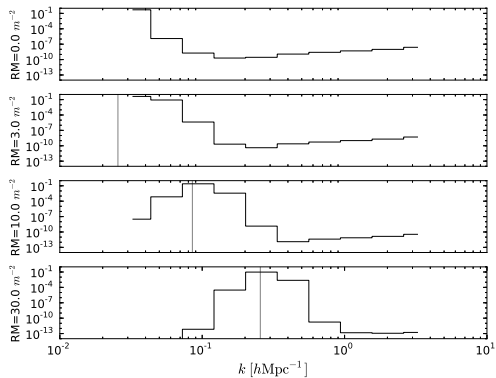}
    \caption{\label{fig:cosmo_1src} Power spectra of the visibilities plotted in Figure \ref{fig:rm_spec}, computed for the median redshift bin of the PAPER band ($z\sim 8$). This plot demonstrates the coupling between rotation measure and $k$-modes, and confirms that Faraday-rotated spectra scatter power to higher $k$ modes than would be contained within the horizon. Grey, vertical lines show the maximally-infected $k$-mode, predicted by equation \ref{eq:k_infected}. The rise in power at high $k$ is due to the $\sim 10^{-9}$ sidelobe of the Blackman-Harris filter. The units of the y-axis are tuned so the total power integrates to one.}
\end{figure}

\begin{figure}
	\plotone{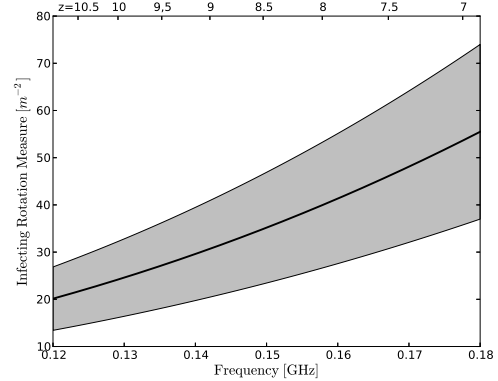}
	\caption{\label{fig:Badguys} The rotation measure most affecting $k\approx 0.15\ h\text{Mpc}^{-1}$, calculated from equation \ref{eq:k_infected}. Shaded region indicates the range of rotation measures affecting $ 0.1 \lesssim k \lesssim 0.2\ h\text{Mpc}^{-1}$.}
\end{figure}

Figure \ref{fig:bh_pspec} shows the Fourier transform over frequency of the spectra in Figure \ref{fig:rm_spec}. While
this doesn't exactly represent the delay-spectrum of a visibility --- there is no beam-weighting, and no $\exp\{-2\pi i
\vec{b}\cdot\hat{s}\}$ component, which essentially defines the delay spectrum --- we interpret it as the delay-structure introduced by a polarized source behind a
Faraday screen. The results of these transforms over the full simulated band are shown in figure \ref{fig:cosmo_1src}.  The most important feature of this plot is this: there is a $k$ mode associated with each rotation measure at each redshift. We can construct an analytic estimate of this $k$ mode by setting the argument of the exponents of a delay mode and a rotation measure mode to sum to zero. First, we approximate the cosmological $k$-mode sampled by a delay mode as $\tau \approx k_{||}dr_{||}/d\nu$. Next, we recall the cosmological scaling from frequency into $h\mathrm{Mpc}^{-1}$,
\begin{equation}
 \frac{dr_{||}}{d\nu} = \frac{dr_{||}}{dz}\frac{dz}{d\nu} = -\frac{c(1+z)}{H(z)\nu}.
 \label{eq:Y}
\end{equation}
Finally, we set $k_{||}\cdot dr_{||}/d\nu\cdot \nu + 2 RM \lambda^2 = 0$. Substituting Equation \ref{eq:Y} for the derivative, we derive an expression for the $k$-mode most affected by a rotation measure $RM$:
\begin{equation}
    	k_{||} \approx \frac{2}{c}  \frac{H(z)}{1+z} \cdot RM \lambda^2,
	\label{eq:k_infected}
\end{equation}
where $H(z)$ is the Hubble parameter. Figure \ref{fig:Badguys} shows a plot of the most culpable rotation measure versus frequency and redshift.

\subsection{Full-Sky Simulation}\label{sec:sim}

To better grasp the effects of Faraday leakage into the 21cm signal, we generate several random realizations of the sky,
each consisting of many polarized point sources. Each source passes through a Faraday screen with some rotation measure,
chosen from a distribution based on current measurements. Next, we simulate that source for a single baseline. Finally, we calculate the power spectrum measured by that visibility. Only one visibility needs to be simulated, because the delay-spectrum approach makes use of the fact that each baseline measures the 21cm EoR with a range of $k$-modes determined by the baseline length, orientation, and bandwidth.

As mentioned before, we use PAPER as a model for choosing instrumental parameters. Our simulated band goes from 120 to 180 MHz, with 60 kHz channels. We use a simulated primary beam model of the PAPER dipole, which has a full width at half max of around $45^\circ$ at 150 MHz \citep{Pober2012}.

Rather than creating an exact simulation of the physical sky, we create a simulation whose statistical properties are physically motivated. This choice reflects a desire for simple, easily tunable parameters for the simulation. In that same spirit, we model all sources simply as point sources with a Poisson distribution. The simulation's primary concern with the spectral information of polarized foregrounds allows us to justify neglecting the angular terms. This is equivalent to assuming for all the relevant $k$ modes, $k_{||} \gg k_{\perp}$. Emphasis on the $k_{||}$ or spectral  modes also motivates our decision to model the sky as numerous point sources. For a more detailed discussion of these effects,  we direct the reader to \citet{Jelic2010}. 

Source positions are distributed uniformly over the sphere. A single source's altitude $\theta$ is drawn from a distribution in which $\cos\theta$ is uniform on $[0,1]$. A source's azimuthal angle $\phi$ is drawn independently from $\cos\theta$ from a distribution uniform on $[0,2\pi]$. This choice of source position distributions conserves the density per area of sources across the sky, and is equivalent to drawing both direction cosines, $(l,m) = (\sin\theta\cos\phi, \sin\theta\sin\phi)$, from a uniform distribution on $[-1,1]$.

In order to achieve realistic source fluxes and source counts, we base the distributions from which we draw various parameters on previous radio surveys. For the source fluxes, we aim to agree with VLSS \citep{VLSS}, NVSS surveys \citep{NVSS}, and the 6C survey \citep{Hales1988}. For the polarization information, we aim to agree with polarized measurements taken by the NVSS survey, particularly, their nearly full sky rotation measure map \citep{Taylor2009}, as well as the polarized survey of the VLSS. 

We present two scenarios for the source counts. First, we draw from the 6C distribution \citep{Hales1988}, taken in the PAPER band at 151MHz. Second, we extrapolate VLSS source counts \citep{VLSS} from 74 MHz to 150 MHz, using a spectral index of -0.79,  following the work of \citet{Cohen2004}. The latter source counts provide more sources at higher flux, which infect the power spectrum from the $I$ visibilities, as we will show in the following sections. Both number counts are consistent with the recent measurements by the Murchison Widefield Array \citep{Williams2012}, an instrument similar in many regards to PAPER.

The differential number counts ($dN/dS$) found by \citet{Hales1988} may be characterized by two power laws, turning over at some knee flux $S_o$
\begin{equation}
    \frac{dN}{dS} = \begin{cases} 4000\ S_o^{-0.76}S^{-1.75}\ Jy^{-1}sr^{-1} 
                        & S_{min} \le S < S_o \\
                                  4000\ S^{-2.81}\ Jy^{-1}sr^{-1} 
                        & S_o \le S \end{cases}
    \label{eq:dNdS}
\end{equation}

Following the 6C survey, we choose the turning point, $S_o$ to be 0.88 Jy. The number of sources simulated (14,855)  is chosen by the size of the PAPER beam at 151 MHz (0.76 sr) and a flux range over which to integrate. We choose to include those sources in between 100 mJy and 10 Jy. This choice provides a reasonable dynamic range of sources. Below the lower limit, the 6C sources are unreliable, and we assume sources above 10 Jy may be easily identified and removed.

If we were to blindly extrapolate the 6C source counts below the lower limit of the catalog, we would add a negligible amount of power. Integrating $S^2 dN/dS$ down to some minimum flux estimates the contribution of the sources above that flux to the total variance of flux. By inserting the 6C source counts, we find that we are including $\sim70\%$ of the estimated total variance. Extending the minimum flux would add more power to the simulation, but would not drastically alter the results of this paper.

VLSS source counts follow a single power law, given by \citet{Cohen2004} as  
\begin{equation}
    \frac{dN}{dS} = 4865\ S^{-2.3}\ Jy^{-1}sr^{-1}. 
\end{equation}

For these source counts, we choose minimum and maximum fluxes of 0.8 Jy and 100 Jy, respectively, rejecting sources well below the lower limit of the catalogue, and providing a reasonable dynamic range for the included sources. Integrating over the PAPER beam provides 6995 sources. These source counts are not qualitatively different from the 6C counts at low fluxes, but they do differ substantially in that they provide many more bright sources.

For both source counts, we also assign a spectral index to each individual source spectrum, drawn from a normal
distribution, mean -0.8, standard deviation 0.1, which roughly agrees with the findings of \citet{Helmboldt2008}.

Each source is also assigned a random polarization angle $\chi$, uniformly sampled from $[0,\pi]$. Total flux is multiplied by a polarization fraction, chosen to reflect the studies of \citet{Tucci2012}. We sample the polarized fraction ($\Pi$) from a log-normal distribution whose mean is 2.01\% and whose standard deviation is 3.84\%. Because the log-normal distribution is not upper-bounded, we reject any drawing over 30\%. Following the aforementioned study, we do not impose any correlation between source flux and polarization fraction. It has been noted that, among other effects, bandwidth depolarization causes the polarized fraction to decrease at lower frequencies \citep{Law2011}. This along with measurements from \citet{Pen2009} indicates that these  distributions, taken at 1.4 GHz, may overestimate the distribution at 150 MHz. We neglect these effects, taking the 1.4 GHz distribution at face value, since the mean polarization fraction can be thought of as a scale factor to the overall power spectrum. 

We base our distribution of rotation measures on the map presented in \citet{Oppermann2012}. To mimic the effects of
depolarization due to a finite spatial resolution \citep[e.g.,]{Law2011}, we apply a low-pass filter to the $RM$ map.
We project the map into a spherical-harmonic basis, and keep only those modes below the resolution of our simulated
instrument. In the case of this simulation, we choose to keep only $\ell \le 100 = 2\pi |u|$. This averages the
polarization vectors in much the same way as a synthesized beam, and it's effect is to essentially remove outliers in
the $RM$ distribution, to which instruments like PAPER may not be sensitive. We then randomly draw rotation measures from the computed cumulative distribution function of $RM$'s given in the \citet{Oppermann2012} data. Aside from low-pass filtering, no spatial information from the data is used. Section \ref{sec:corr} briefly discusses the negligible consequences of spatially correlating $RM$.

\begin{figure}
   \plotone{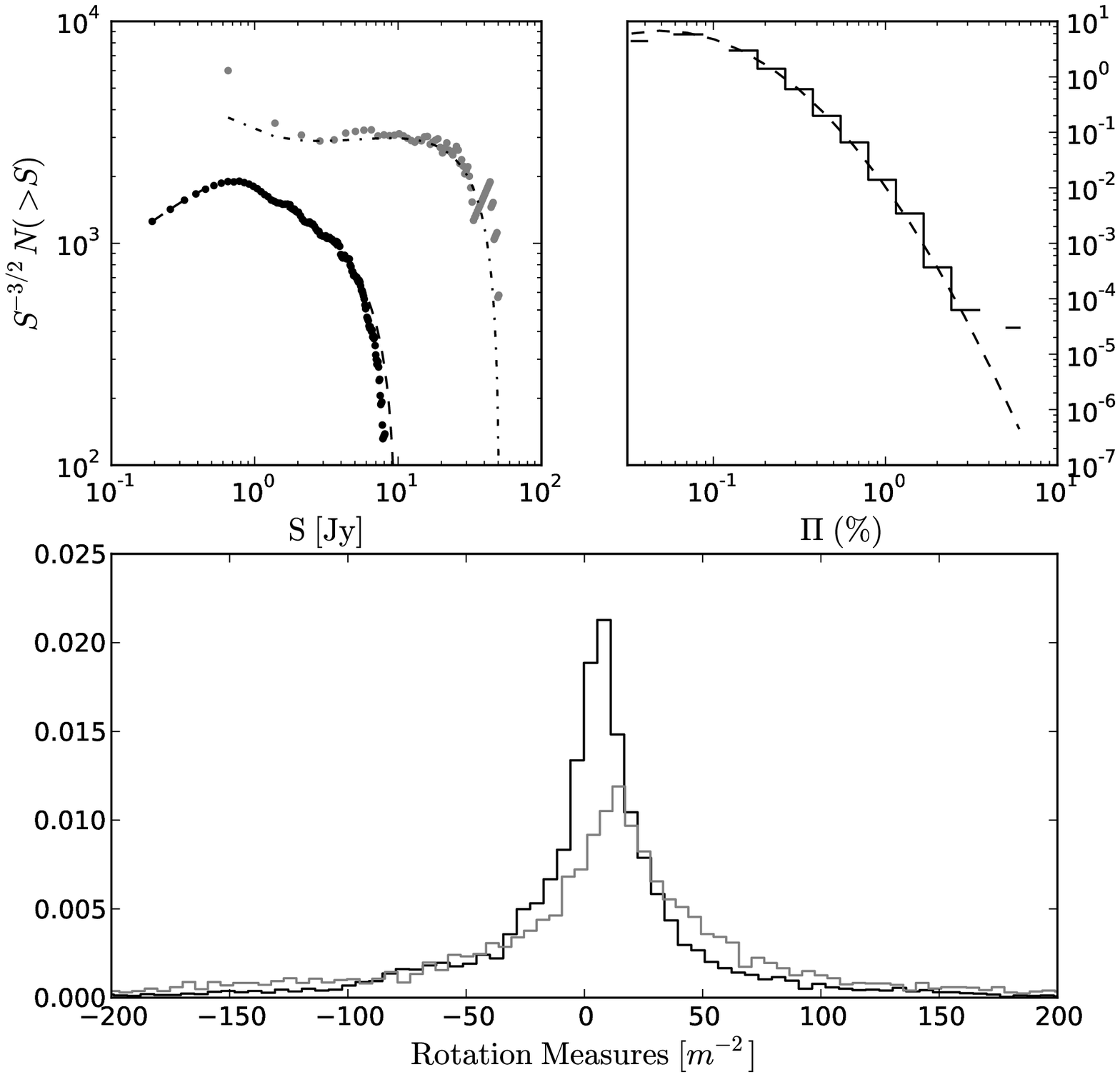}
   \caption{\label{fig:SimParams} Distributions of simulated parameters. (Top Left) $S^{-3/2}N(>S)$ source counts of
   unpolarized flux. In black, the 6C source counts, from \citet{Hales1988}, and in grey, the VLSS counts, from
   \citet{VLSS}. (Top Right) Normalized histogram of the log-normal distribution of polarized fraction, taken from
   \citet{Tucci2012}. (Bottom) Normalized histograms of the two distributions of rotation measures. Black is taken from
   \citet{Oppermann2012}, grey is that with a doubled standard deviation. The $RM$ distribution extends to several
   hundred m$^2$, but we restrict the extent of the $x$-axis to highlight the distribution, rather than the width of its
   tails. In the upper two panels, the dashed line is the distribution from which sources are drawn, and the bins or points are the values of one realization of the simulation.}
\end{figure}

Histograms of the distributions of rotation measure, polarized fraction, and source counts can be found in figure \ref{fig:SimParams}. Over-plotted on all is the distribution from which they are drawn. 

We model all sources as point sources, neglecting the effects of any diffuse emission. This choice reflects the desire for a simple, easily tunable parameters in the simulation. While diffuse emission certainly is present, its spectral structure is qualitatively the same as that of a point source, so the results in the frequency direction will not change. Were we to add a diffuse model to the simulation, it would widen a source's response in delay-space, preventing it from being localized. This will not affect the line-of-sight power spectrum, at high $k$, since the spectrally-smooth, diffuse emission still falls within the horizon.

Table \ref{tab:SimLabels} summarizes the three treatments of the simulation we will be using. Simulation A, with 6C
source counts and the Oppermann $RM$ distribution is likely the most accurate. Simulation B steepens the source counts, providing fewer, brighter sources, and simulation C doubles the width of the $RM$ distribution. 

\begin{table}
 \centering
  \begin{tabular}{ c  c  c  c }
	\hline
	Name & Source Counts & $N_{src}$ & $RM$ distribution \\
	\hline
	A & 6C & 14,855 & Oppermann  \\
	\hline
    B & VLSS & 6,995 & Oppermann \\
	\hline
	C & 6C & 14,855 & $2\times$Oppermann \\
	\hline
  \end{tabular}
 \caption{\label{tab:SimLabels} Simulation Treatments}
\end{table}

\begin{figure}
    \plotone{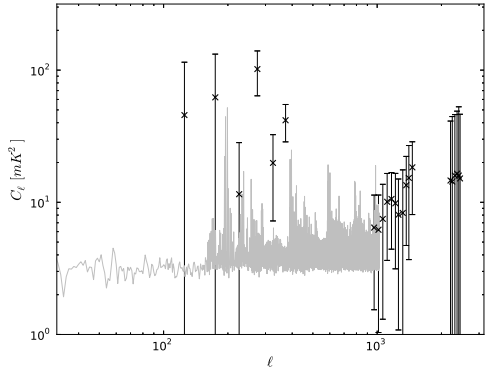}
    \caption{\label{fig:l2Cl} $\ell$ vs. $C_\ell$ for one realization of the simulation (solid line), alongside the results from \citet{Bernardi2010}, (crosses). This roughly demonstrates the agreement of the simulation presented in section \ref{sec:sim} with current observations. The simulation is consistently lower than the observed values due to the lack of extended structure.}
\end{figure}

To check if the results of this simulation are consistent with the measurements in \citet{Bernardi2010}, we compare the
two-dimensional $C_\ell$ power spectrum with that presented in their paper. Figure \ref{fig:l2Cl} shows the 2D power
spectrum of one realization of the simulation, with the constraints from figure 20 of \citet{Bernardi2010} plotted over
it. We see qualitatively that our simulation well obeys the upper-limit imposed by the Bernardi measurement, and proceed with the results.

We calculate the visibilities for a 32m, east-west baseline. This choice reflects the most common spacing of the maximum-redundancy configuration presented in \citet{PAPERSensitivity}. The choice of baseline orientation is arbitrary, and since we are modeling only point sources, the choice of baseline length will only set the horizon limit of the power spectrum. Since the delay affected by a rotation measure is independent of a choice of baseline (equation \ref{eq:k_infected}), choosing a relatively short baseline will isolate the foregrounds at lower $\tau$, and highlight the Faraday leakage. 

\begin{figure}
    \plotone{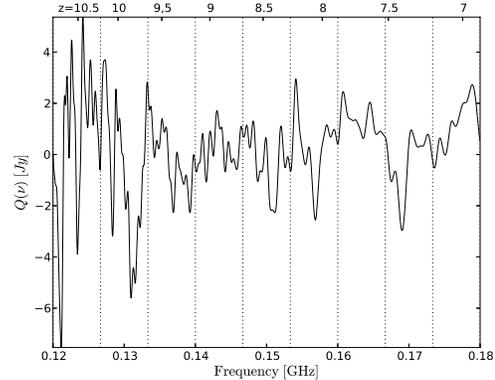}
    \caption{\label{fig:sample_vis} Simulated $Q$ visibility with the parameters shown in figure \ref{fig:SimParams}.}. 
\end{figure}

The full measurement equation used in this simulation is
\begin{align}
    \V(u,v,\nu) = \sum_{j=1}^{N_{src}} A(l_j,m_j,\nu) \Pi_j 
        S_j^{150} \left(\frac{150\ \text{MHz}}{\nu}\right) ^{\alpha_j} \nonumber \\ 
        \times\exp\left\{ - i [2\pi \nu(ul_j + vm_j) + 2RM_j\lambda^2 + 2\chi_j] \right\},
    \label{eq:measurement}
\end{align}
where each source $j$ is assigned a flux ($S_j$), polarization fraction ($\Pi_j$), spectral index ($\alpha_j$), a position ($l_j,m_j$). rotation measure ($RM$), polarization angle ($\chi_j$), and is weighted by the model primary beam ($A$). A sample $Q$ visibility is shown in Figure \ref{fig:sample_vis}.

We choose not to include the parallactic rotation of $Q$ into $U$, implying that the $Q$ we label in this paper are fixed to topocentric, azimuth and altitude coordinates. This choice clarifies equations and allows for an ease of understanding which would be obfuscated by writing both $Q$ and $U$.

\subsection{Simulated Power Spectra}\label{sec:sim_pspec}

\begin{figure*}
    \plotone{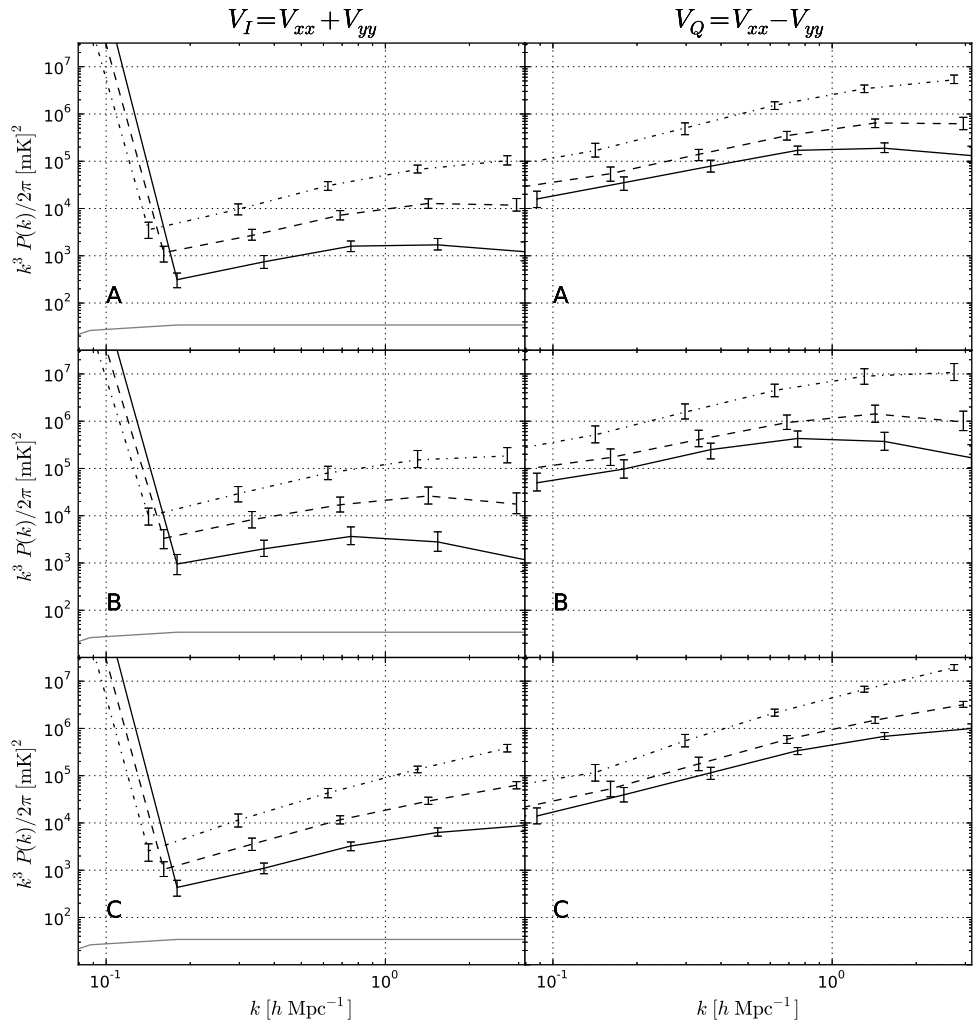}
    \caption{\label{fig:sim_pspec} Power spectral measurements for the three treatments of the simulation shown in Table
    \ref{tab:SimLabels}. The left column shows the power spectra of the $I$ visibilities, and the right shows the
    spectra of $Q$ visibilities. The top row shows simulation treatment A, the second row shows treatment B, and the
    third row shows treatment C. Line styles depict different redshift bins: 7.25 (solid), 8.33 (dashed), 9.73 (dot-dashed).  Grey line give a toy model of the expected 21cm emission from \citet{FOB}. Both
    visibilities include contributions from both the (intrinsic) Stokes $I$ and $Q$. The simulated levels of $Q$
    emission indicate that polarized leakage into $I$ needs to be less than for or five orders of magnitude in mK$^2$.}
\end{figure*}

Figure \ref{fig:sim_pspec} show the power spectra for source counts from the extrapolated VLSS and 6C surveys, as well as the spectrum of 6C source counts with a widened rotation measure distribution. We interpret the power spectrum of the $I$ visibility as the amount of polarized leakage corrupting the EoR signal (henceforth called $Q \to I$ leakage), and the $Q$ visibility's power spectrum is our best representation of the polarized signal. These plots show the median power in each $k$ bin for 1000 realizations of the simulation, with error bars show the one-sigma extent of the bandpowers for these realizations. These power spectra confirm the prediction of section \ref{sec:single_source},  that $\lambda^2$ phase wrapping extends the foreground cutoff presented in \citet{DelaySpectrum} to higher $\tau$ bins, corrupting some of the most sensitive regions of $k$ space for 21cm EoR analysis. They also demonstrate the prediction in that section that high-redshift bins will be most affected.

\begin{figure}
    \plotone{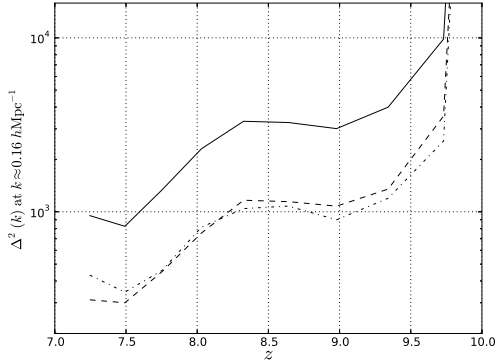}
    \caption{\label{fig:k3pk_v_z} Median $Q\to I$ leakage for the $k\approx 0.16 h\text{Mpc}^{-1}$ bin of the power spectrum, $\Delta^2(k)$ vs. $z$ for all treatments of the simulation shown in figure \ref{fig:sim_pspec}, and defined in table \ref{tab:SimLabels}: A (solid), B (dashed), C (dot-dashed). We note that with every treatment, the amount of polarized flux leaking into the power spectrum grows with $z$, confirming that leakage is worse at lower frequencies.}
\end{figure}

The severity of the leakage can be inferred from the power in the most EoR-sensitive $k$ bins ($0.2\ h\text{Mpc}^{-1} \lesssim k \lesssim 0.3\ h\text{Mpc}^{-1}$). Figure \ref{fig:k3pk_v_z} shows $\Delta^2(k)$ in these bins as a function of redshift. The leaked power ranges in the thousands of $\text{mK}^2$, increasing from high frequency / low redshift to low frequency / high redshift. These estimates are about two orders of magnitude above level of the expected 21cm signal \citep{Lidz2008}. If we take this simulation as an accurate prediction of the low-frequency sky's polarized emission, these results imply that na\"{i}vely adding $\V_{xx}$ and $\V_{yy}$, formed with an approximately 10\% asymmetric primary beam, incorporates enough bias from polarized leakage to completely obscure the 21cm signal. The levels of leakage in our simulations demands a strategy to model and remove polarized sources.

We note that simply adding $\V_{xx}$ and $\V_{yy}$ will also remove a negligible component of the  EoR signal via the same mechanism. In a sense, $Q\to I$ leakage can be thought of as a rotation of power between the two Stokes parameters. Hence, for precision measurements of the EoR signal, this simple estimate may not be ideal. However, the effect of $I\to Q$ leakage is small (compare the levels of $\V_{Q}$ and high-$k$ modes of $\V_{I}$) and should not provide a significant hindrance to detection. 

\subsection{Correlated Polarization Vectors}\label{sec:corr}

The results of the previous section were intended to extrapolate previous measurements to low fluxes and investigate the
effects of an unresolved forest of dim, polarized point sources. It neglects the known spatial correlations of the rotation measure distribution
\citep{Kronberg2011}. Furthermore, the random drawing of polarization angle could have a cancelling effect on the
visibilities. This neglect could potentially suppress our estimation of polarized leakage into the power spectrum.

\begin{figure}
    \plotone{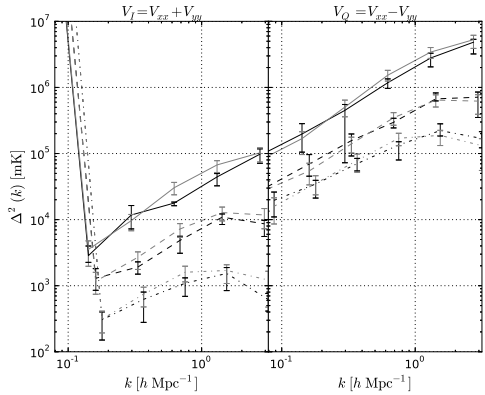}
    \caption{\label{fig:CorrPspec} A comparison of power spectral measurements for a treatment of the simulation with correlated polarization angles
        (black), and those with treatment A from Table \ref{tab:SimLabels} (gray). As in figure \ref{fig:sim_pspec}, the left-hand
        panel shows the power spectrum derived from the $I$-visibility, and the right-hand panel, the $Q$-visibility. Three
        redshift-bins are included, and denoted by linestyles: 9.73 (solid), 8.33 (dashed), and 7.25 (dot-dashed). The results
        of simulation A agree with the results of correlating polarization vectors, which indicates that our choice of
        randomly-assigning polarization angles and rotation measures is valid.}
\end{figure}

To investigate the possible effects of correlating the polarization vector, we include a treatment of the simulation
where we choose rotation measures from the Oppermann map \citep{Oppermann2012}, with a pointing center at the Galactic
south pole --- a reasonable field for EoR analysis. We then set all polarization angles to zero, maximally correlating
polarization vectors, while still including information of the polarized sky. All other simulation parameters are
identical to simulation A of Table \ref{tab:SimLabels}. Figure \ref{fig:CorrPspec} compares the
results of this treatment with simulation A of the previous section. The power spectrum of this treatment agrees with
simulation A at all redshifts and values of $k$, for both polarizations. This agreement indicates that spatial
correlations in $RM$ and polarization angle do not significantly affect polarized leakage into the power spectrum. Thus
the assumption of the previous section that the polarization vectors are spatially uncorrelated does not affect the
results of this paper.

\section{Mitigation}
\label{sec:remove}

Section \ref{sec:SIM} predicts an excess polarized signal due only to point sources at around $10^4\ \text{mK}^2$ at
$k\sim 0.15$ for most treatments of the simulation. While the exact levels of these predictions may be subject to some
error, the need certainly arises for some removal scheme, which must roughly suppress power from polarized foregrounds
by around four orders of magnitude in the power spectrum.

\begin{figure}
	\plotone{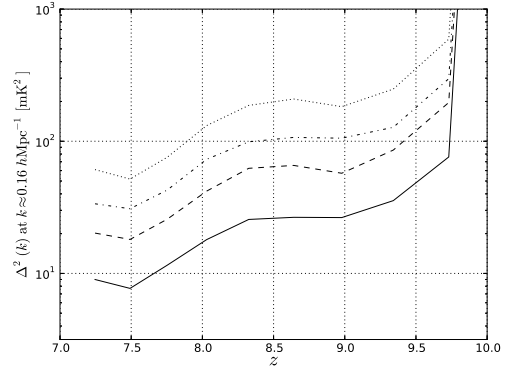}
	\caption{\label{fig:k3pk_v_Nrm} $\Delta^2(k)$ vs $z$ at a single $k \approx 0.16\ h\mathrm{Mpc}^{-1}$. Each line style shows the volume-averaged power after the brightest $N$ sources were removed from the simulation. The number of removed sources, $N$ is depicted by line style. Dotted: 1000 sources removed (40 mJy), dot-dashed; 2000 removed (25 mJy), dashed: 3000 removed (18 mJy), and solid: 5000 removed (11 mJy).}
\end{figure}

To investigate the effects of modelling and removing polarized sources, we rerun the simulation, excluding the brightest polarized sources. Figure \ref{fig:k3pk_v_Nrm} shows the median value of 500 simulations of one $k$-bin of the power spectrum for removing the brightest 1000, 2000, 3000, and 5000 sources. These limits in number of sources correspond to flux-limits of 40 mJy, 25 mJy, 18 mJy, and 11 mJy, respectively. We remove these sources from treatment A of the simulation, which includes around fifteen thousand sources. Despite having removed nearly one-third of the sources, the leaked power still exceeds 10 mK$^2$. 

To remove enough flux to consistently fall below the expected EoR signal, we needed to remove eight to ten thousand
sources ($\sim$6 mJy), roughly the number of sources in the simulation. We recomputed these fluxes with a lower minimum
flux (60 mJy), expecting a similar result, but found that we increased the power in this $k$-bin by one or two mK$^2$.
Ten to twelve thousand sources required removal for the total power to fall below 10 mK$^2$. We exclude further
investigation of this analysis for three reasons. First, current measurements do not constrain $dN/dS$ to the levels
necessary to accurately model such low-flux sources. Second, including lower-flux sources does not significantly affect
the result that the expected polarized power spectrum will be of the order of $10^4$-$10^6$ mK$^2$. Third, the variance
from one simulation to the next in the power was large enough that the two treatments of the simulation --- even with ten thousand sources removed --- could not be considered significantly different.

These onerous levels of source-removal suggest that a different mitigation scheme be considered. Future instruments may take polarization into consideration in design. Leakage can be mitigated with more circular beams, and circular feeds avoid $Q\to I$ leakage entirely. Even with existing data, rotation measure synthesis \citep{BrentjensDeBruyn} could potentially provide the power to separate sources with distinct $RM$ structure to be separated from EoR signal. 

\section{Discussion}

Were a power spectrum computed from only one linearly polarized visibility ($xx$, for instance), all polarized power
would corrupt the measurement. We have chosen to suppress the polarized leakage by adding linearly-polarized
visibilities. The leakage is dependent on the difference of the two beams, and by having beams that are at most ten
percent different suppresses the signal by around two to three orders of magnitude. Correcting for the beam-weighting in the image domain can
further suppress the leakage, but errors in the beam model will introduce leakage in much the same manner. Hence, the
constraint of having to suppress polarized leakage by four orders of magnitude causes the need for an accurate primary beam model to
around the one percent level in the case of imaging, or symmetric at the one percent level if visibilities are used
directly. 

These estimates of power are also dependent on the relative strengths of diffuse, polarized emission and polarized point
sources. We have taken care to agree with current measurements, but we note that above $\ell \sim 300$, the current
constraints are noisy. We have interpreted them as an upper limit. In the limiting case where diffuse emission is the
only component to the polarized sky, this leakage could be suppressed by measuring with a longer baseline, which in turn
measures lower $\ell$ and $k_\perp$. We have chosen to use a sixteen-wavelength baseline, which corresponds to $\ell
\approx 200$. This choice of baseline length is relatively short for interferometers at these wavelengths, but falls at
the high end of the \citet{Bernardi2010} detection. 

Including diffuse emission in the simulation would certainly increase the total power in the simulation for low $\ell$,
but the frequency structure would remain qualitatively the same as point sources.  As we showed in Section \ref{sec:corr}, the
correlation of rotation measure and polarization angle that could be introduced by an extended structure does not
significantly affect the power spectrum. For this reason, we can consider the polarized sky as having two components
with nearly identical footprints in the power spectrum: diffuse and point like. Both components will exhibit similar
frequency structure, so choice of baseline length will set the relative weightings of these components. \citet{Bernardi2010} briefly discuss
some of the implications of their measurement of extended structure to the three-dimensional power spectrum in their
conclusion, which agrees with our analysis of point-like structure.

We conclude our discussion of the simulation results by noting the large variance in the simulated power. The results shown are the median band-powers in $\Delta^2(k)$ for 500 realizations of the simulation. Taking so many realizations into account essentially maps out the posterior distribution of the $\Delta^2(k)$ bandpowers. The one-sigma width covers nearly an order of magnitude, which indicates the level of Faraday leakage is highly sensitive to the exact parameters drawn in any realization. The actual level of leakage measured will thus be highly dependent on a choice of field, and on cosmic variance. 

\section{Conclusion}
We have predicted the three-dimensional power spectrum of polarized emission around 150 MHz to be in the range of
$10^4$-$10^6$ mK$^2$ at $k_{||} \sim 0.15\ h\text{Mpc}^{-1}$. These predictions were based on simulations movitvated by
current observations of the polarized sky at 150 MHz and 1.4 GHz. An elliptical beam provides one mechanism for this
power to leak into a measurement of the unpolarized signal. Using a fiducial model of the PAPER beam, we estimated this leakage to be in the thousands of mK$^2$, several orders of magnitude above the expected 21cm EoR signal. Modelling and removing polarized sources may eliminate much of this leakage, but these simulations suggest the amount of removal required far exceeds reasonable capabilities of current instruments.

Work is currently underway to measure frequency structure of polarized power, to investigate the amount of its leakage into the $I$ power spectrum, and to better characterize the polarized radio sky using existing PAPER data.

\hfill

This analysis and the PAPER project are supported through the NSF-AST program (award 1129258). Computing resources were provided by a grant from Mt. Cuba Astronomical Foundation.

\end{document}